# Shaping liquids into space structures

## Microgravity-assisted design and manufacturing of minimal surfaces

*Erez Hochman[1], Aaron Sprecher[2], Amos A. Hari[3], Moran Bercovici[4]*
*[1,2,3]Technion – The Israel Institute of Technology [4]ETH Zurich*
*[1]erez.hochman@campus.technion.ac.il*
*[3]a.hari@campus.technion.ac.il*
*[2]asprecher@technion.ac.il*
*[4]moran.bercovici@mat.ethz.ch*

**Abstract**

This work advances a fundamentally new approach to space-based construction by using liquid self-organization in microgravity as a generative design and fabrication principle. Building on the *LiquiFab* method, we demonstrate how minimal-surface architectures - traditionally dependent on complex additive manufacturing - can instead emerge directly from the physics of fluid interfaces shaped by programmable boundary conditions. We present a simulation-to-fabrication workflow, that includes a boundary-driven minimal-surface solver integrated in Grasshopper/Rhino, and an experimental system that implements in a neutral buoyancy environment simulating microgravity. This enables the generation of customizable Schwarz-P-inspired with tunable geometry and thickness, illustrating a scalable pathway for material-efficient, on-orbit fabrication. To validate performance in true microgravity, a flight experiment on the International Space Station is scheduled for the first quarter of 2027 and we here detail the additional considerations required for this experiment.

**Keywords**

form-finding, space-oriented design, in-space manufacturing, LiquiFab

**INTRODUCTION**

As human activity in space expands, there is an increasing demand for in-situ and on-demand construction methods capable of operating directly in space (Arney et al., 2021). Recent efforts have focused on adapting terrestrial manufacturing technologies for use aboard the International Space Station (ISS), including the introduction of in-situ 3D printing in 2014 (Prater & Werkheiser, 2018). While additive manufacturing offers geometric flexibility and on-demand production, it remains constrained by the printer's build envelope and by the requirement for the printer to visit each voxel in space, which limit scalability.

A recent technique, LiquiFab (Hochman et al., 2025), demonstrated a new fabrication approach: a volume of liquid polymer is injected into an immiscible immersion liquid matched in density, so that buoyancy precisely cancels gravity and surface tension becomes the sole active force. In this effectively weightless state, a small set of solid boundary surfaces placed in the fluid drives the liquid away from its natural spherical form and into a new minimum-energy configuration that conforms to the prescribed constraints. Once the desired shape is reached, UV exposure solidifies the polymer into a solid object.

That work also introduced the concept of "LiquiBricks" - modular units assembled sequentially to build up larger constructs. Once a LiquiBrick is solidified, its faces and edges serve as the boundary conditions for the next injection: fresh liquid polymer is introduced, wets the existing solid surfaces, and adopts a new minimum-energy form continuous with the preceding element. Repeating this stepwise process, complex three-dimensional assemblies can be built without a conventional mold or print head. Theoretically, the process is scale-invariant: in the absence of gravity the LiquiFab approach could be implemented at any size. In practice, however, the achievable brick size in the Earth-based neutral-buoyancy setting is bounded by the precision of density matching, the curing kinetics of the polymer, and the capacity for heat dissipation during the exothermic polymerization reaction. In Space, the first limitation is eliminated, but the remaining two persist.

Building on this foundation, the present study focuses on the design and realization of minimal-surface LiquiBricks inspired by the Schwarz-P unit cell, a triply periodic minimal surface (Schoen, 1970), presented in Figure 1. Minimal surface lattice structures are widely recognized in aerospace engineering for their high strength-to-weight ratio and efficient material distribution and are therefore ideal candidates for construction of space structures. However, due to their geometric complexity their fabrication typically relies on additive manufacturing. In this work we show that the *LiquiFab* approach can readily produce such structures. We introduce a programmable design-to-fabrication workflow of customized thin-shelled minimal surface geometries and the development of a semi-automated experimental system operating in a simulated microgravity environment. The experiment objective is to create a physical element within the bounding box with the dimensions of: 80x80x80 $mm^3$. We design the experimental system such that it could be directly adapted for demonstration on the International Space Station, planned for the first quarter of 2027.

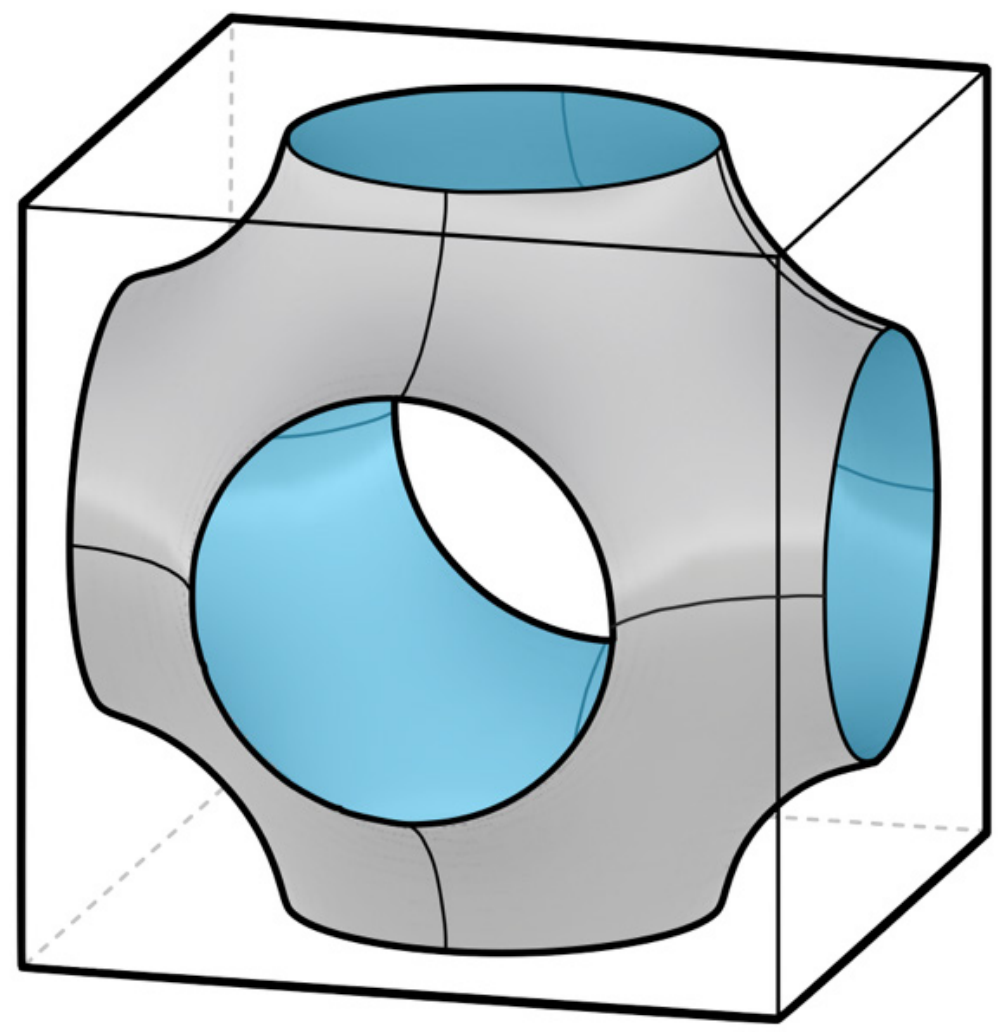

Figure 1: A Schwarz-P surface is a primitive triply periodic minimal surface, defined by zero mean curvature and periodicity in three spatial directions. Mirroring the basic unit in three planes creates the presented object, which we refer to as a unit cell. Interestingly, the unit cell intersects with a bounding cube at three circles - a property that we will leverage in this work for construction of the object by *LiquiFab.*

## SIMULATION TOOL

The Schwarz-P surface is a mathematical minimal surface of zero thickness (Figure 1). Any physical realization possesses finite volume and is therefore an approximation of the ideal geometry. *LiquiFab* treats fabrication as a constrained minimum-energy problem: the liquid interface satisfies prescribed boundaries, while the remaining surface relaxes to its minimal-energy configuration.

To approximate the Schwarz P topology, we define six annular rings placed on the faces of a cube (Figure 2b). These rings act as fixed boundary curves defining the geometric constraints for the minimum-energy problem. To compute the shape of the object, we use Surface Evolver (Brakke, 1992), which we have integrated into the Rhino/Grasshopper environment.

The user defines the cube dimensions together with lists of the inner and outer radii of the rings that are to be located at each face (Figure 2a). Based on these inputs, our Grasshopper script generates the geometric entities required by Surface Evolver, including the boundary curves and surfaces and an initial watertight mesh (Figures 2b–e). Once the initial mesh is constructed, it undergoes a remeshing procedure intended to standardize element quality and accelerate convergence. Surface Evolver then iteratively adjusts vertex positions to minimize the total surface energy while preserving the prescribed boundary constraints and the initial volume enclosed by the mesh, unless a different volume is specifically defined by the user. Throughout this process, the geometry progressively evolves from the initial coarse mesh toward a stable minimal-energy configuration (Figures 2f–i). After convergence, a final mesh refinement step is applied to achieve the desired resolution and geometric fidelity, yielding the final minimal-surface approximation shown in Figure 2j.

This workflow provides the computational basis for the design variations and fabrication studies presented in the following sections, linking minimal-surface theory with fluid-interface physics and boundary-driven geometric control.

**Code availability**

https://github.com/Fluidic-Technologies-Laboratory/surface_evolver_grasshopper

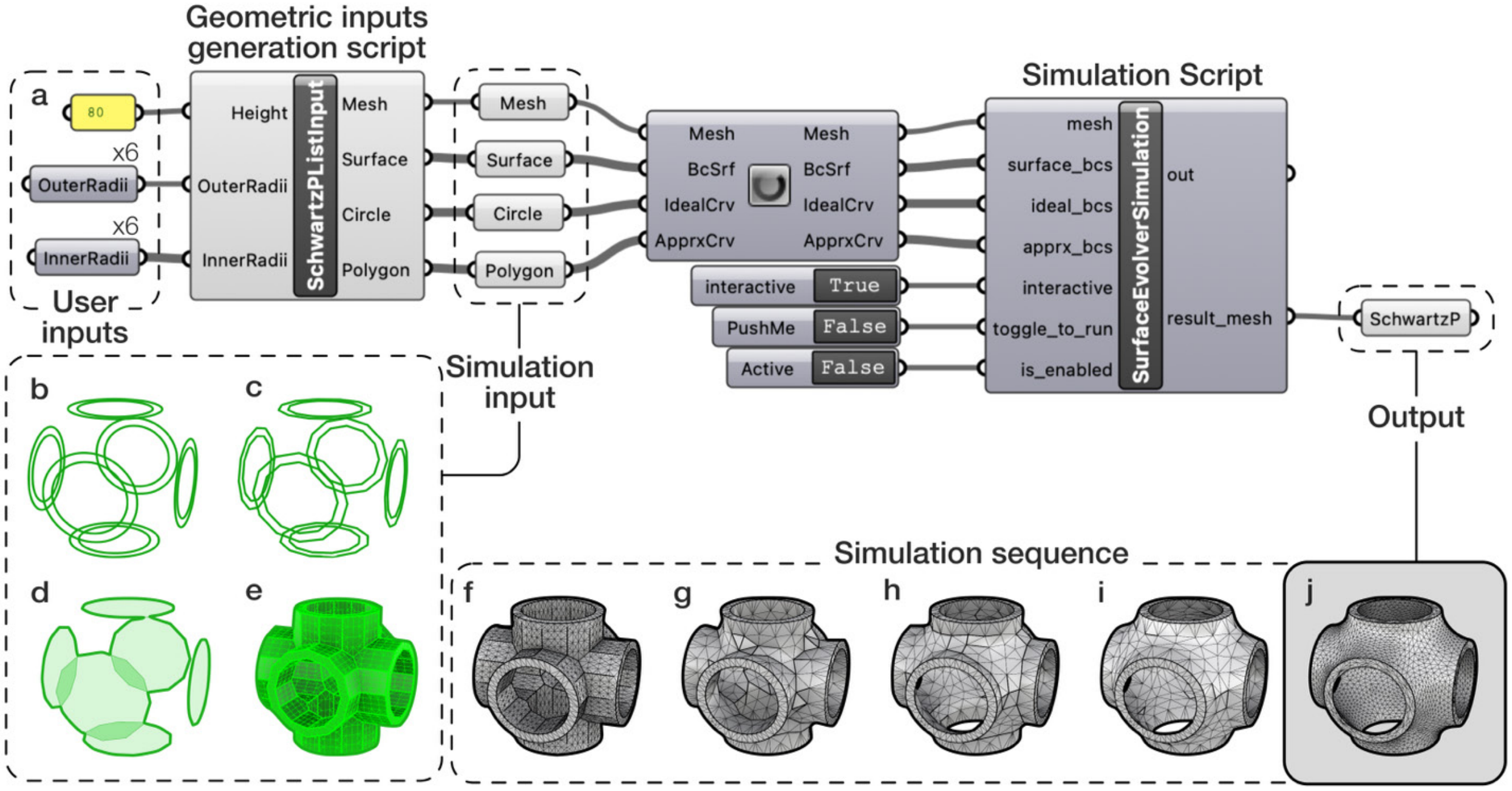

Figure 2: Simulation script workflow in the Rhino Grasshopper environment. (a) Control parameters defining the relative locations and the dimensions of the boundaries. (b-e) Our Grasshopper script generates the geometric entities (circles, polygons, surfaces and mesh) required as input to Surface Evolver. (f-i) Simulation steps showing the gradual convergence of the liquid volume to its minimum energy state (j) Final mesh refinement followed by further convergence to a fine resolution.

## DESIGN FOR SPACE APPLICATIONS

The design of structures intended for use in space must balance functional, geometric, and material considerations. Habitable architectures, in particular, require openings for circulation, interfaces for docking or assembly, and the capacity for routing equipment while maintaining a highly material-efficient envelope. Since all liquid material must be transported to orbit, minimizing the volume needed for construction is essential. We envision a space habitat, whose envelope is constructed entirely in space from the assembly of individual LiquiBricks, each on the scale of meters to tens of meters.

The classical Schwarz-P unit cell provides an appealing starting point for such applications: it encloses space efficiently with minimal surface area and connecting several units in series naturally forms interconnected passages. However, defining the geometry through boundary elements - rather than prescribing the minimal surface explicitly - opens up a richer design space. This boundary-driven approach allows the creation of Schwarz-P-inspired surfaces that preserve the beneficial characteristics of minimal surfaces while relaxing strict triply periodic symmetry. In this section we use our computational model to demonstrate how variations in boundary conditions yield diverse LiquiBrick units that can be assembled into larger architectural systems.

In addition to boundary geometry, we also investigate the role of the *liquid volume* used during fabrication. Adjusting this parameter directly influences how material distributes itself within the resulting shell - yielding structures that range from thin, lightweight envelopes to thicker, more robust walls - while still conforming to the minimal-energy configuration imposed by the boundaries. Together, these parameters define a coherent and versatile design-to-fabrication strategy for generating space-relevant minimal-surface structures.

**Boundary elements**

Figure 3 illustrates how the geometry of a *LiquiBrick* can be systematically shaped by manipulating the boundary conditions that define it. The sequence begins with a reference configuration (Figure 3a), consisting of six annular rings arranged symmetrically on the faces of a cube. This base form provides a controlled starting point from which several derivative geometries are produced.

**Variation in ring diameter.** The first family of variations (Figure 3b) explores the effect of increasing the nominal diameter of the boundary rings while keeping their positions fixed. Enlarging these diameters changes two essential characteristics. First, the resulting openings between adjacent rings become wider, which is crucial in space-structure applications where internal circulation of air, transfer of equipment, cable routing, or crew access must be accommodated. Second, increasing the ring diameter reduces the amount of liquid required to “close the surface” - that is, the volume of polymer needed to create a continuous minimal-energy interface spanning all boundary rings.

**Variation in ring thickness.** The second set of variations (Figure 3c) adjusts the thickness of the boundary rings - that is, the width of the annular face to which the liquid surface attaches. This face can be made narrow c(i) or wide c(ii), providing a simple geometric control over the extent of the contact region. Ring thickness therefore governs how each unit interfaces with adjacent components, enabling the design of broader docking areas, reinforced attachment zones, or more compact connection points.

**Combined and customized configurations.** The third series (Figure 3d) demonstrates how variations in diameter and thickness can be combined to create highly customized boundary configurations. These examples show that even non-uniform or asymmetric rings reliably generate coherent minimal-surface geometries. This flexibility enables the encoding of functional distinctions directly into the boundary layout.

**Assembly of multiple units.** Figure 3e demonstrates that individual LiquiBricks can be combined into larger assemblies by aligning their boundary configurations. Different unit types - such as those derived from figure 3d(i) and figure 3d(iv) - can be arranged to create structures that accommodate various functional requirements in a modular way. Here, the continuity of openings, alignment of interface regions, and compatibility of local curvatures all emerge naturally from the boundary-driven design process.

Together, the variations illustrated in figure 3 demonstrate a versatile design space in which inter-module compatibility arises from deliberate control of boundary geometry.

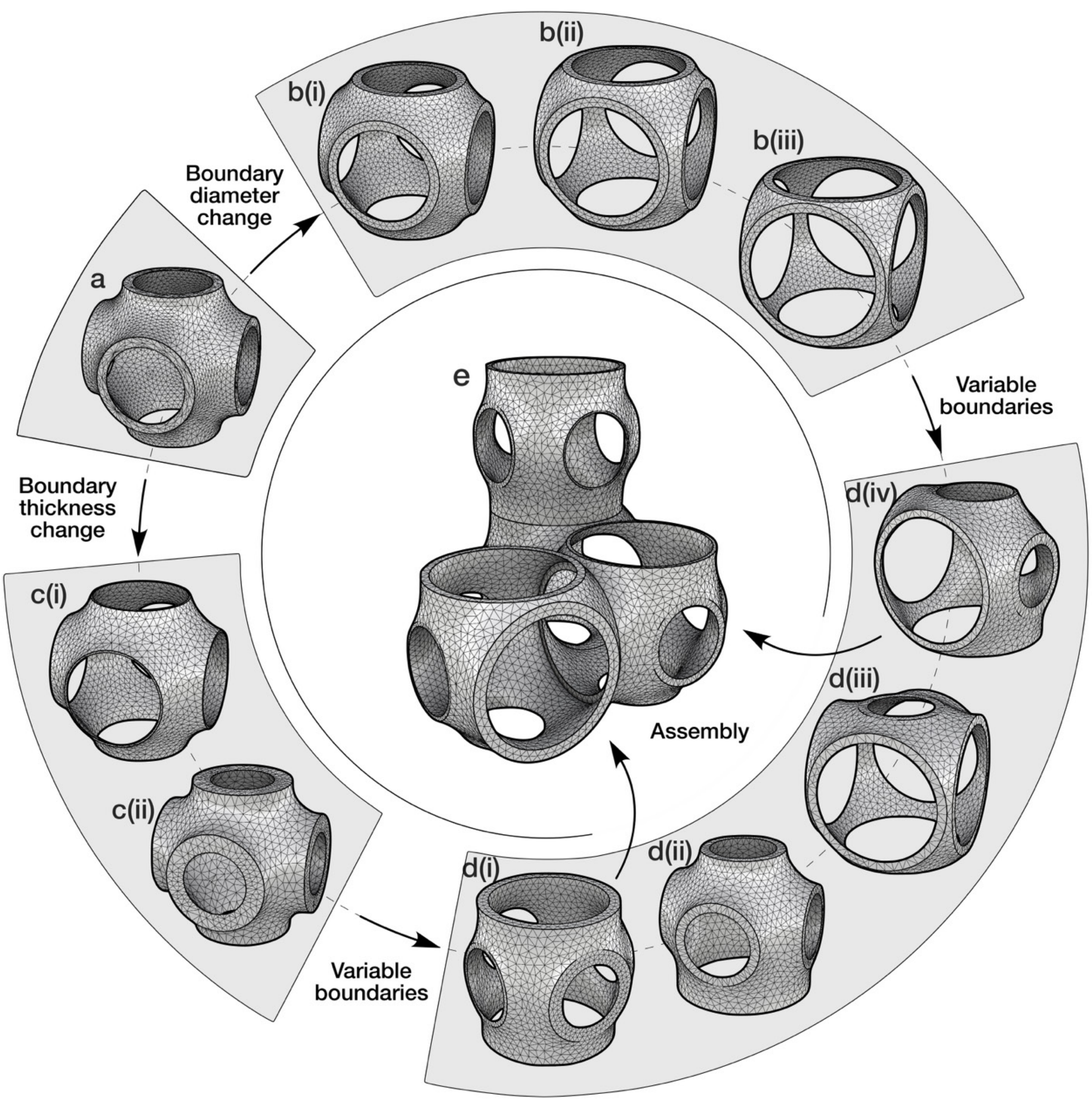

Figure 3: Boundary-driven variations of a *LiquiBrick*. (a) Reference configuration of six annular rings. (b) Diameter variations showing larger openings and reduced volume needed to form a continuous minimal-surface interface. (c) Thickness variations controlling the width of the surface–boundary contact region. (d) Combined diameter-and-thickness configurations producing diverse unit geometries. (e) Assembly of multiple units into larger structures through aligned boundary configurations.

## Liquid volume

The enclosed volume is an additional degree of freedom in shaping a LiquiBrick. Figure 4 illustrates how changes in this parameter affect the distribution of wall thickness across a unit. Figures 4a–d show a representative geometry sectioned along a diagonal plane to expose its internal structure.

Figures 4e–h then present a sequence of units generated from identical boundary conditions but with different prescribed volumes. Smaller volumes produce geometries with pronounced variations in local thickness, including regions where the inner and outer surfaces come into close proximity. At a specific volume threshold, the surface converges toward a more uniform thickness. Increasing the volume beyond this point leads to an overall thicker shell. Across all cases, the thickness at the boundary interfaces remains fixed by the predefined boundary geometry.

These examples demonstrate that enclosed volume provides a straightforward means of adjusting the global shell characteristics while preserving the minimal-surface topology imposed by the boundary conditions.

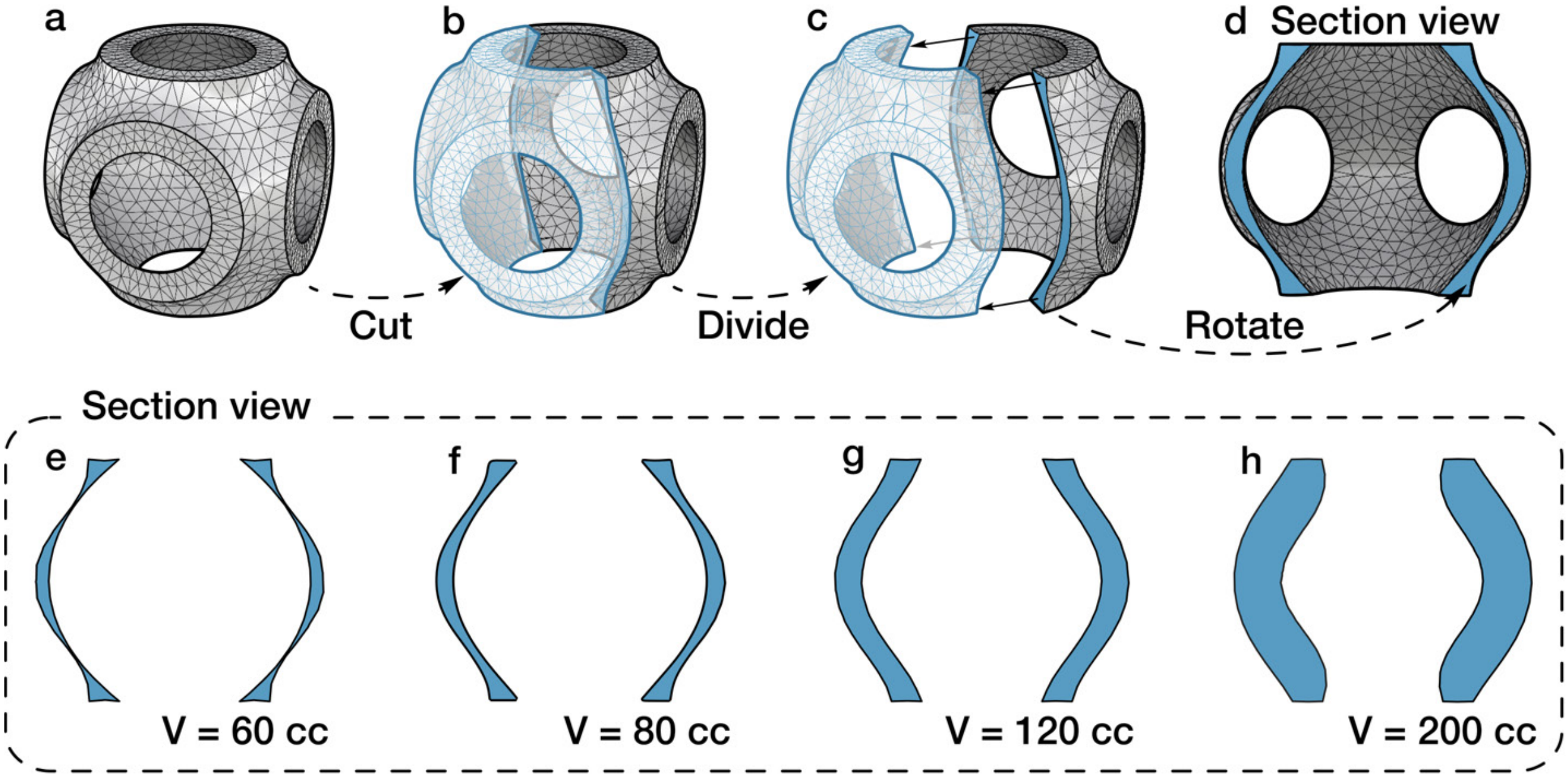

**Figure 4:** Influence of liquid volume on the wall-thickness distribution of a LiquiBrick. (a–d) Sectioning of a representative geometry along a diagonal plane to reveal its internal structure. (e-h) Units generated from identical boundary conditions with different liquid volumes, showing how small volumes create areas where surfaces approach each other, intermediate volumes yield a more uniform thickness, and larger volumes produce thicker shells. In all cases, the boundary-interface thickness is fixed by the ring geometry.

## LAB EXPERIMENT

The objective of this experiment was to demonstrate the realization of physical minimal-surface structure inspired by a Schwarz-P element. It examined the tooling, fabrication logic and steps required to achieve such element, with particular emphasis on the role of boundary conditions as primary design drivers.

To simulate microgravity on Earth, we carried out the experiment in a neutral-buoyancy tank using Plateau's density-matching method (Plateau, 1857). We adjusted a water-glycerol solution to match the density of a curable photopolymer, allowing the polymer to behave as if weightless within the fluid. Next, we placed a Fixator containing six annular boundary rings into the tank and introduced a controlled volume of liquid photopolymer through its boundary inlet, initiating the formation of a suspended polymeric blob (Figure 5b). We then inflated the blob using an immersion-liquid inflation device, which expanded the polymer until it contacted all boundary surfaces (Figure 5c–d). Once it had fully expanded, we used a jetting nozzle to puncture the polymeric membrane enclosed within the boundary rings, equalizing the internal and external pressures and allowing the interface to relax freely into its minimal-surface configuration (Figure 5e–f). After the surface had relaxed, we removed the sealing plug at the illumination port and inserted a UV illumination device into the center of the form to cure the polymer and solidify the final geometry (Figure 5g).

Figure 6 presents an evaluation of the computational model through a quantitative comparison with the fabricated prototype, scanned using a 3D scanner (GOM Scan 1, Carl Zeiss). Deviations between the two external surfaces were quantified using the GOM Inspect software, revealing a maximum absolute error of 2.75 mm for a model with overall dimensions of 80 × 80 × 80 mm.

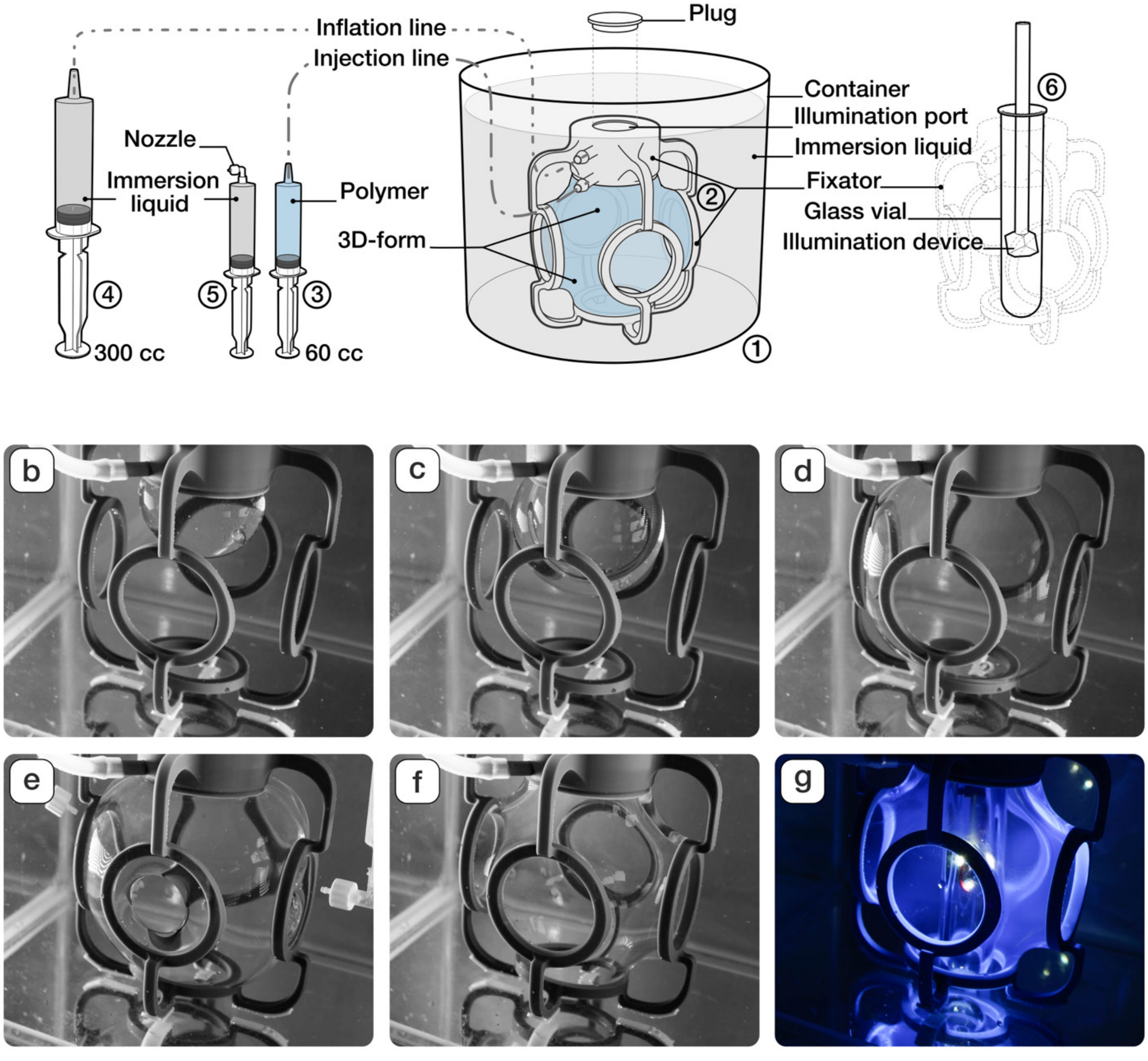

**Figure 5:** Fabrication sequence in a neutral-buoyancy environment. (a) Schematic illustration of the system's components. (b) A Fixator holding six annular boundary rings is placed into the density-matched immersion liquid that simulates microgravity. Liquid photopolymer is injected through the Fixator's inlet, forming an initial suspended blob. (c–d) The blob is inflated with an immersion-liquid, causing it to expand outward until it contacts all boundary surfaces. (e–f) A jetting nozzle punctures the polymeric membrane enclosed within the rings, allowing the interface to relax into its minimal-surface shape. (g) UV illumination cures the polymer and solidifies the object.

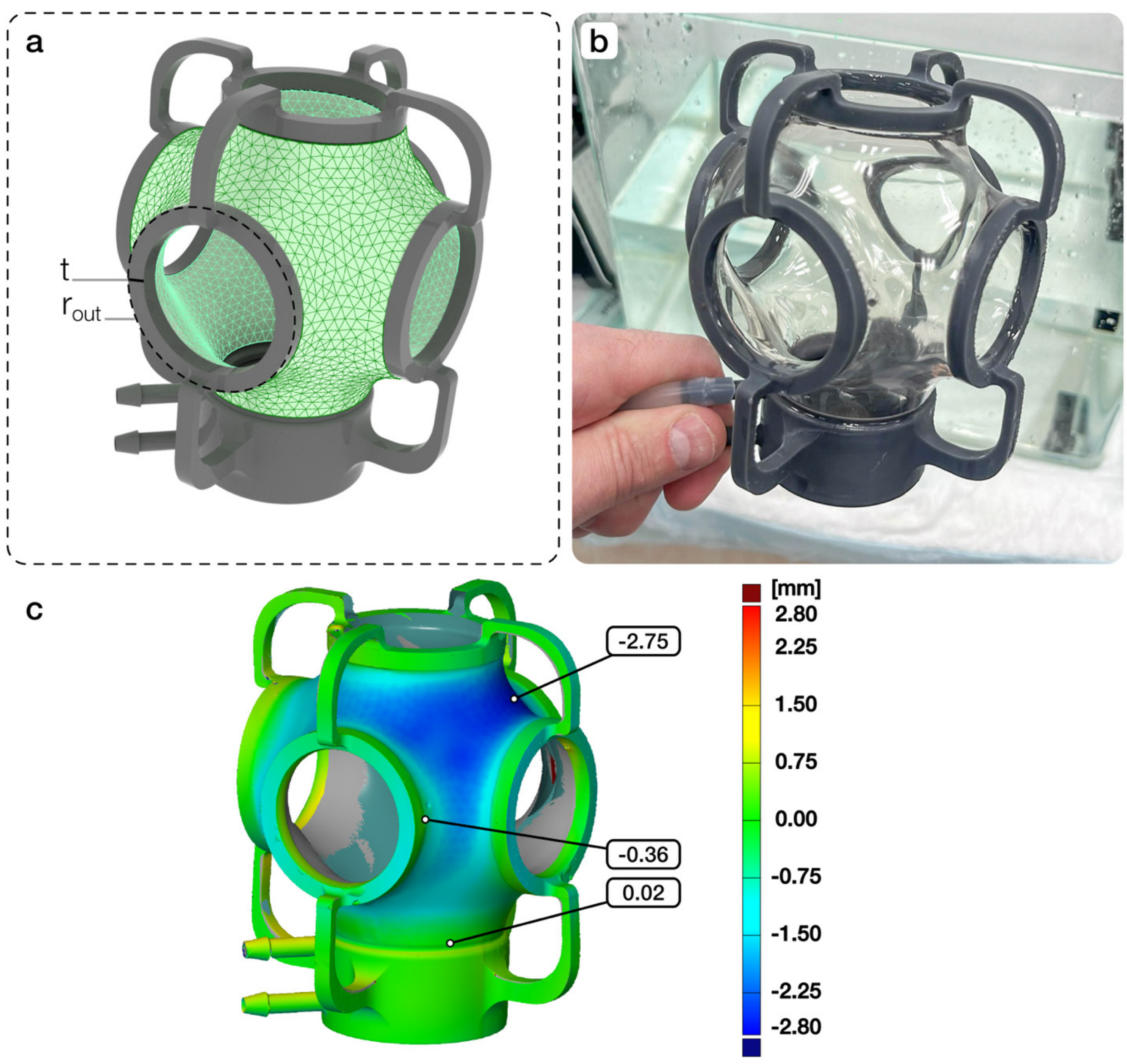

Figure 6: Comparison between the computational reference model and the fabricated Schwarz-P-inspired minimal-surface prototype. The boundary parameters are $r_{out} = 25$ mm and $t = 5$ mm, and the nominal polymer volume was 60 ml, with a $\pm 1$ ml accuracy in the experiment. (a) Computational model. (b) Fabricated prototype. (c) Geometric comparison between the external surface of the 3D scanned fabricated prototype and of the computational reference model. Specific values indicate the maximum deviations along the surface, as well as the values at the pinning points.

## DISCUSSION AND CONSIDERATIONS FOR IMPLEMNTATION ON THE ISS

This paper presents a design-to-fabrication workflow for generating minimal-surface structures through fluid-interface self-organization in microgravity. By integrating simulation, parametric design, and experimental validation, the research demonstrates how boundary conditions, influencing a physical process, can be used to drive form generation.

To date, our experiments were conducted in a neutral buoyancy environment, simulating microgravity conditions. A true microgravity environment presents new challenges, but at the same time also eliminates some of the constraints of neutral buoyancy. As we have learned from prior in-space manufacturing experiments, the precise behavior of a complex system in space cannot always be precisely anticipated. (Ericson et al., 2026) To learn more about the behavior of our system in space, we are planning an experiment on board the international space station (currently scheduled for the first quarter of 2027) in which we intend to produce a Schwarz-P object in space and return the fabricated objects back to Earth for further analysis.

One constraint that is relaxed in space is the choice of polymer. While on Earth the polymer must be insoluble in water and within a specific density range, operation in true microgravity allows the use of essentially any polymer. While the experiments we presented in this paper were performed using a simple polyurethane that cannot be considered as a structural material in space, other polymer such as Dur™ 173S (Grossman et al., 2024) have been shown to be compatible with the conditions in space, and can be directly used with the LiquiFab approach.

While the immersion liquid provides conditions in which the total body force on the liquid is zero, just as in microgravity, the significant liquid mass around the object likely offers additional stabilization that may not exist in space. Furthermore, while on Earth the polymer is inflated with immersion liquid, in space it will be inflated with gas. And finally - the film puncturing dynamics may be significantly different in the absence of an immersion liquid. In other words, while the steady-state shape should be identical on Earth and in space, liquid dynamics may be quite different and can be experimentally investigated only in space.

An additional difference between Earth and Space experiments relates to thermal management. On Earth, free convection, particularly in the presence of the immersion liquid, plays a critical role in dissipating heat generated during the exothermic polymerization process. In microgravity, however, the absence of buoyancy-driven convection limits passive heat transfer. As a result, active cooling (e.g. by forced convection) must be implemented to maintain stable material properties.

Successful implementation of the method in space would validate the feasibility of the approach. Since the method is inherently scale-invariant, i.e. the physical behavior and resulting shapes are independent of size, extrapolation to large scale is straightforward. Naturally, additional design and engineering research would be required for such manufacturing at a large scale.

## References

Arney, D., Sutherland, R., Mulvaney, J., Steinkoenig, D., Stockdale, C., & Farley, M. (2021). *On-orbit Servicing, Assembly, and Manufacturing (OSAM) State of Play, 2021 Edition* (No. 20210022660). NASA.

Brakke, K. A. (1992). The Surface Evolver. *Experimental Mathematics*, *1*(2), 141–165. https://doi.org/10.1080/10586458.1992.10504253

Ericson, J., Widerker, D., Stibbe, E., Elgarisi, M., Katzman, Y., Luria, O., Gommed, K., Razin, A., Hari, A. A., Gabay, I., Frumkin, V., Hamad, H. A., Segal, E., Amouyal, Y., Szobody, T., Ticknor, R., Balaban, E., & Bercovici, M. (2026). *Modeling the Thermal Behavior of Photopolymers for In-Space Fabrication*.

Grossman, E., Atar, N., Bolker, A., Riggs, B. E., Minton, T. K., Gouzman, I., & Vidavsky, Y. (2024). Space Durable 3D Printed High-Performance Polymers Based on Cyanate Ester/Extended-Bismaleimide. *Advanced Functional Materials*, *34*(30), 2313255. https://doi.org/10.1002/adfm.202313255

Hochman, E., Sprecher, A., Suzina, K., Mann, A., Mihalovich, Y., Frumkin, V., & Bercovici, M. (2025). *LiquiFab—Building with liquids in weightlessness* (arXiv:2512.19756). arXiv. https://doi.org/10.48550/arXiv.2512.19756

Plateau, J. (1857). I. Experimental and theoretical researches on the figures of equilibrium of a liquid mass withdrawn from the action of gravity.–Third series. *The London, Edinburgh, and Dublin Philosophical Magazine and Journal of Science*, *14*(90), 1–22.

Prater, T. J., & Werkheiser, N. J. (2018). *Summary Report on Phase I and Phase II Results From the 3D Printing in Zero-G Technology Demonstration Mission, Volume II*.

Schoen, A. H. (1970). *Infinite periodic minimal surfaces without self-intersections* (Vol. 5541). National Aeronautics and Space Administration.